\documentstyle[12pt,epsf]{article}

\newlength{\extraspace}
\newlength{\extraspaces}
\catcode`\@=11
 
\def\numberbysection{\@addtoreset{equation}{section}
\def\theequation{\arabic{section}.\arabic{equation}}}

\begin{document}
%
\thispagestyle{empty}
\begin{center}
\begin{flushright}
TIT/HEP--332 \\
{\tt hep-th/9606157} \\
June, 1996 \\
\end{flushright}
\vspace{3mm}
\begin{center}
{\Large
{\bf Vacuum Structures of Supersymmetric Yang-Mills Theories in $1+1$ 
Dimensions 
}} 
\\[18mm]

{\sc Hodaka Oda},\footnote{
\tt e-mail: hoda@th.phys.titech.ac.jp} \hspace{2.5mm}
{\sc Norisuke Sakai}\footnote{
\tt e-mail: nsakai@th.phys.titech.ac.jp} \hspace{2.5mm}
and \hspace{2.5mm}
{\sc Tadakatsu Sakai}\footnote{
\tt e-mail: tsakai@th.phys.titech.ac.jp} \\[3mm]
{\it Department of Physics, Tokyo Institute of Technology \\[2mm]
Oh-okayama, Meguro, Tokyo 152, Japan} \\[4mm]

\end{center}
\vspace{18mm}
{\bf Abstract}\\[5mm]
{\parbox{13cm}{\hspace{5mm}
Vacuum structures of supersymmetric (SUSY) Yang-Mills theories 
in $1+1$ dimensions are studied with the spatial direction 
compactified. 
SUSY allows only periodic boundary conditions for both fermions 
and bosons. 
By using the Born-Oppenheimer approximation for the weak coupling 
limit, we find that the vacuum energy vanishes, and hence the SUSY 
is unbroken. 
Other boundary conditions are also studied, 
especially the antiperiodic boundary condition for fermions 
which is related to the system in finite temperatures. 
In that case we find for gaugino bilinears a nonvanishing vacuum 
condensation which indicates instanton contributions. 
}}
\end{center}
\vfill
\newpage
\vfill
\newpage
\addtocounter{section}{1}
\setcounter{equation}{0}
\setcounter{section}{0}
\setcounter{footnote}{0}
\numberbysection




\vspace{7mm}
\pagebreak[3]
\addtocounter{section}{1}
\setcounter{equation}{0}
\setcounter{subsection}{0}
\setcounter{footnote}{0}
\begin{center}
{\large {\bf \thesection. Introduction }}
\end{center}
\nopagebreak
\medskip
\nopagebreak
\hspace{3mm}


Supersymmetric (SUSY) theories offer promising models for the unified 
theory, 
but nonperturbative methods are acutely needed to make further progress 
in understanding such issues as supersymmetry breaking. 
Recently, 
good progress has been made on nonperturbative 
aspects of supersymmetric Yang-Mills theories. Using holomorphy and 
duality, 
exact results of the low energy physics of $N=2$ super Yang-Mills theories 
were obtained \cite{SW}. As for $N=1$ super Yang-Mills theories, further 
insight has been gained within the context of duality \cite{Sei;Ku}. 
These results deal with low energy effective theories, and are based on 
general but rather indirect arguments. 
It is perhaps illuminating to study the supersymmetric gauge theories 
by more dynamical calculations. 
Since it is still very difficult to study $4$-dimensional gauge theories, 
we would like to start from $1+1$ dimensions. 
In $1+1$ dimensions, gauge fields have no dynamical 
degrees of freedom. 
If matter fields belong to the fundamental 
representation of a gauge group, 
they become tractable in the $1/N$ approximation, and provide an 
illuminating model for confining theories \cite{THooft}. 
If there are matter fields in adjoint representations, the $1/N$ 
approximation is not sufficient to solve the $SU(N)$ gauge theories. 
Still a number of recent studies have been performed both numerically 
and analytically for Yang-Mills theories with adjoint matter fields, 
and the studies have yielded several interesting results 
\cite{DaKl}-\cite{KoZh}. 
The Born-Oppenheimer approximation in the weak coupling region 
has been used to study the vacuum structure of gauge theories 
with adjoint fermions \cite{Lenz}. 
Since the gauge coupling in $1+1$ dimensions has the dimension of 
mass, the weak coupling is characterized by 
\begin{equation}
gL \ll 1 ,
\end{equation}
where $L$ is the interval of the compactified spatial dimension. 
The fermion bilinear was found to possess a nonvanishing vacuum 
expectation value which exhibits instanton-like dependence 
on gauge coupling. 
The Yang-Mills theories with adjoint 
fermions were also studied at finite temperature and 
were shown to be dominated by instanton effects at high temperatures 
\cite{Smilga}. 
The Born-Oppenheimer approximation has been used to study SUSY
gauge theories in four dimensions \cite{Wit}\cite{Sm}.

Since the gauge fields have no dynamical degree of freedom, 
SUSY Yang-Mills theories in $1+1$ dimensions
(SYM$_2$)
contain both spinor field (gaugino) 
and scalar field 
in the adjoint representation. 
A manifestly supersymmetric (infrared) regularization scheme has been obtained 
recently using the discretized light-cone approach 
\cite{MSS}. 
In this study, numerical results suggested an exponentially rising 
density of states. 
Our understanding of these theories is, however, not yet 
sufficient. In particular, vacuum structures such as 
the vacuum condensate need to be clarified. 
We need alternative systematic 
approaches to study them thoroughly, 
since the light-cone approach which is best suited to deal 
with excited states is notoriously laborious when applied to the unraveling 
of vacuum structures. 
The concept of zero modes is crucial in understanding vacuum structures 
\cite{MaYa}, \cite{KoSaSa}. 
As for the possibility of the SUSY breaking, the Witten index 
of the SUSY Yang-Mills theories 
has been calculated recently, and was found to be nonvanishing \cite{Li}. 
Although this result implies no possibility for spontaneous SUSY 
breaking, we feel it still worthwhile to study the vacuum of the SUSY 
Yang-Mills theories in $1+1$ dimensions by a more detailed dynamical 
calculation, since the calculation of the Witten index involved a certain 
regularization of bosonic zero modes which may not be easily 
justified. 

The purpose of this paper is to study vacuum structures of supersymmetric 
Yang-Mills theories in 1+1 dimensions. 
We use the Born-Oppenheimer approximation in the weak coupling 
region, as used for non-SUSY Yang-Mills theories with adjoint fermions 
\cite{Lenz}. 
To formulate the weak coupling limit, we need to 
compactify the spatial direction. 
Since gauge fields naturally follow periodic boundary conditions, 
we need to require the same periodic boundary conditions for scalar 
and spinor fields in order not to break SUSY by hand. 
We have found that the ground state has a vanishing vacuum energy, 
suggesting that SUSY is not broken spontaneously. 
This result is consistent with the result on the Witten index \cite{Li}. 
We also examine all four possibilities of periodic and anti-periodic 
boundary conditions for fermions and bosons. 
The case involving antiperiodic fermions and periodic bosons 
should be related to the case of finite temperature. 
We find in this case that the vacuum energy does not vanish, 
and the gaugino bilinear exhibits nonvanishing vacuum condensation. 
The vacuum condensate turns out to have nontrivial dependence 
on the dimensionless constant $gL$, which 
resembles the instanton contributions. 
It would be interesting to conduct a study to see if 
our results can be 
explained by instanton contributions.


Sec. II briefly summarizes the Hamiltonian approach to SUSY 
Yang-Mills theories in $1+1$ dimensions.
The canonical quantization is carried out,
and dynamical degrees of freedom are identified. 
In Sec. III, vacuum structures of SYM$_2$ are discussed using 
an adiabatic approximation.
Sec. IV discusses cases involving antiperiodic boundary conditions 
for spinors or scalars, 
and Sec. V contains a summary. 

%
%
%
%
%
\vspace{7mm}
\pagebreak[3]
\addtocounter{section}{1}
\setcounter{equation}{0}
\setcounter{subsection}{0}
\addtocounter{footnote}{0}
\begin{center}
{\large {\bf \thesection. 
SUSY Yang-Mills Theories 
in $1+1$ Dimensions}}
\end{center}
\nopagebreak
\medskip
\nopagebreak
\hspace{3mm}

%
%

Since gauge fields have no dynamical
degrees of freedom in two dimensions,
the SUSY gauge multiplet in $1+1$ dimensions 
consists of gauge fields $A^{\mu}$,
the Majorana spinor $\Psi$, and the real scalar $\phi$ 
\cite{Ferrara}. 
SUSY $SU(N)$ Yang-Mills action is given 
by $\Psi$ and $\phi$ fields in the adjoint 
representation as 
$\phi \equiv \phi^a t^a$ and $\Psi \equiv \Psi^a t^a$,
where the $t^a$ are generators of $SU(N)$ with the normalization 
${\rm tr}\left( t^a t^b \right ) = \frac{1}{2} \delta^{a b}$
\cite{Ferrara}  
\begin{equation}
{\cal L} = {\rm tr} \left\{- \frac{1}{2} F_{\mu \nu} F^{\mu \nu}
+ D_\mu \phi D^\mu \phi
+ i \bar{\Psi} \gamma^\mu D_\mu \Psi
- i g \phi \bar{\Psi} \gamma_5 \Psi \right\}. 
\end{equation}
The gauge coupling is denoted as $g$, 
$D_\mu = \partial _\mu + i g [A_\mu,~] $ is the usual covariant 
derivative and 
$F_{\mu \nu} = \partial_\mu A_\nu - \partial_\nu A_\mu +
i g [A_\mu,A_\nu]$. 
The action is invariant under the following supersymmetry
transformations \cite{Ferrara} (also see \cite{MSS}).
\begin{equation}
\delta A_\mu = i \bar{\epsilon} \gamma_5 \gamma_\mu \Psi,
\quad
\delta \phi = - \bar\epsilon \Psi , \quad
\delta \Psi = - \frac{1}{2} \epsilon \epsilon^{\mu \nu}
F_{\mu \nu} + i \gamma^\mu \epsilon D_\mu \phi. \quad
(\epsilon^{01}= -\epsilon_{01}=1)
\end{equation}
Taking the following representation of $\gamma$ matrices 
\begin{equation}
\gamma^0 = \sigma_2,\quad
 \gamma^1 = i \sigma_1,\quad 
\gamma^5 \equiv \gamma^0 \gamma^1 = \sigma_3, \quad 
C = -\sigma_2, 
\end{equation}
the Majorana spinor $\psi=C{\bar \psi}^T$ is real.

In this paper we compactify the spatial 
direction to a circle with a finite radius $L/2\pi$. 
The gauge fields naturally follow periodic 
boundary conditions 
\begin{equation}
A^\mu (x=0) = A^\mu (x=L) .
\end{equation}
We shall specify boundary conditions for $\Psi$ and $\phi$ later. 
%

Gauge theories have a large number of redundant 
gauge degrees of freedom which should be eliminated by a gauge-fixing 
condition.
In this paper we quantize the system in the Weyl gauge,
\begin{equation}
A^0 = 0 .
\end{equation}
We can impose Gauss' law as a subsidiary condition for the 
physical state $| \Phi \rangle$ 
\begin{equation}
[ D_1 E^a(x) -g \rho^a(x) ] | \Phi \rangle =0, 
\quad 
\rho^a 
=
 f^{abc} \phi^b \pi^c
+ \frac{i}{2} f^{abc} \Psi_\alpha^c \Psi_\alpha^b.
\label{Gauss cond.} 
\end{equation}
where $\pi^a$ and $-E^a\equiv F^{a01}$ are the conjugate variables 
of $\phi^a$ and $A^{a1}$ respectively, and 
$\rho^a$ is the color charge density, 
and $f^{abc}$ are the structure constants of the Lie algebra of $SU(N)$ : 
 $[t^a, t^b] = i f^{abc} t^c$. 
Note that the Gauss law determines $E$, except for 
its constant modes $e$. 
One can eliminate $A^1$ by using an appropriate gauge transformation, 
except for the $N-1$ spatially constant modes $a^p$ which are given by 
\begin{equation}
{\cal P}\exp \left( i g \int_0^L dx A^1(x) \right) = 
V e^{i g a L} V^\dagger ,\quad (a = a^p t^p),\label{a^p def}
\label{constantmode;a}
\end{equation}
where $V$ is a unitary matrix. 
Hereafter we shall use the convention that 
$a, b, \cdots =1,2 \cdots, N^2-1$ 
represent the indices of the generators of $SU(N)$, 
and $p,q, \cdots =1,2 \cdots, N-1$ represent those of Cartan subalgebra. 
The commutation relation between $a^p$ and 
$e^q$ is given as \cite{LeNaTh} 
\begin{equation}
\left[ e^p , a^q \right] = i \delta^{p q} \qquad
p,q = 1,\ldots,N-1.
\end{equation}

In the physical state space, we can eliminate 
redundant gauge degrees of freedom by solving 
the Gauss law constraint (\ref{Gauss cond.}), and find 
the Hamiltonian 
\begin{eqnarray}
H
&\!\!\!
=
&\!\!\!
\int_0^L dx {\cal H}(x) 
=
K_a + H_{\rm c} + H_{\rm b} + H_{\rm f} + H_{\rm int} , 
\label{hamiltonian;1} \\
K_a 
&\!\!\!
=
&\!\!\!
 \frac{1}{2L} \sum_p e^{p \dagger} e^p,\\
H_{\rm c} &=& \frac{g^2}{L} \sum_{n=- \infty}^{\infty}
\sum_{ij} \int_0^L dy \int_0^L dz 
(1 - \delta_{ij} \delta_{n0} )
\frac{ \left( \rho(y) \right)_{ij} \left( \rho(z) \right)_{ji}}
{\left( \frac{2 \pi n}{L} + g (a_i - a_j) \right)^2}
e^{2 \pi i n (y-z)/L} ,\label{H_c} \\
H_{\rm b} 
&\!\!\!
=
&\!\!\!
 \int_0^L dx \left\{ 
\frac{1}{2} \pi^a \pi^a + 
\frac{1}{2} \left(D_1 \phi \right)^a 
\left( D_1 \phi \right)^a \right\},\\
H_{\rm f} 
&\!\!\!
=
&\!\!\!
 \int_0^L dx \left(- \frac{i}{2} \right)
\Psi^a \sigma_3 \left( D_1 \Psi \right)^a , \qquad
H_{\rm int} 
=
 \int_0^L dx \,
{\rm tr} \left\{ i g \phi \bar{\Psi} \gamma_5 \Psi \right\} 
\label{H_int}
\end{eqnarray}
where $ a_i = a^p t^p_{ii}$  with no summation over $i$ implied, 
and $\sum_i a_i = 0$. 
Here the covariant derivative $D_1$ contains only the zero mode of 
$A^1$ : $D_1 = \partial_1 - i g [a,~]$. 
One should note that gauge fields $A^\mu$, except the zero modes $a^p$, 
are completely eliminated.

In order to investigate the vacuum structures of our model, we solve 
Schr\"odinger's equation with respect to the Hamiltonian 
(\ref{hamiltonian;1}) 
\begin{eqnarray}
H| \Phi \rangle = E |\Phi \rangle ,
\end{eqnarray}
where $|\Phi \rangle$ denote state vectors in the physical space.
Because of hermiticity of the variables $a$, 
the kinetic energy $K_a$ is given in terms of the 
Jacobian $J[a]$ of the transformation (\ref{constantmode;a}) 
\cite{LeNaTh} 
\begin{eqnarray}
K_a 
&\!\!\!
= 
&\!\!\!
\frac{1}{2L} e^{p \dagger} e^p 
= - \frac{1}{2L} \frac{1}{J[a]} \frac{\partial}{\partial a^p}
J[a] \frac{\partial}{\partial a^p}, \\
J[a] 
&\!\!\!
=
&\!\!\!
 \prod_{i>j} \sin^2 
\left( \frac{1}{2} g L (a_i - a_j) \right). 
\label{J}
\end{eqnarray}

In analogy with the radial wavefunctions, it is useful to define 
a modified wave function 
\begin{equation}
\tilde{\Phi}[a] \equiv \sqrt{J[a]}\Phi [a] .
\label{radial}
\end{equation}
The kinetic energy operator for $\tilde \Phi$ is 
(with the notation $\partial_{p} = \partial/\partial a^{p}$), 
\begin{equation}
K_a^\prime \equiv 
\sqrt{J} K_a {1 \over \sqrt{J}} 
= -\frac{1}{2L} \partial_p \partial_p + V^{[N]} ,
\label{K_a^'}
\end{equation}
\begin{equation}
V^{[N]} 
\equiv 
\frac{1}{2L} \frac{1}{\sqrt{J}}\left(\partial_p \partial_p \sqrt{J}
\right)
= - \frac{(gL)^{2}}{48L} N (N^{2}-1) .
\label{V}
\end{equation}
Thus we obtain a boundary condition
for the modified wavefunction,
\begin{eqnarray}
\tilde{\Phi}[a]=0, \, \ \ {\rm if} \quad J[a]=0 \ .
\label{boundary}
\end{eqnarray}


Let us now quantize the fields $\Psi$ and $\phi$. 
The gauge field zero modes $a^p$ couple only to 
off-diagonal elements, which are parameterized as : 
$\varphi_{ij} = \sqrt{2} \Psi_{ij}$, 
$\varphi_{ij}^\dagger = \sqrt{2} \Psi_{ji}$, 
$\xi_{ij} = \sqrt{2} \phi_{ij}$, 
$\xi_{ij}^\dagger = \sqrt{2} \phi_{ji}$, 
$\eta_{ij} = \sqrt{2} \pi_{ij}$, and 
$\eta_{ij}^\dagger = \sqrt{2} \pi_{ji}$
$(i<j)$.
With these conventions the Hamiltonian takes the form
\begin{eqnarray}
H_{\rm f} 
&\!\!\!
=
&\!\!\!
 H_{\rm f,diag} + H_{\rm f,off}, \qquad 
H_{\rm b} = H_{\rm b,diag} + H_{\rm b,off} ,\\
H_{\rm f,diag} 
&\!\!\!
=
&\!\!\!
 \frac{1}{2i} 
\sum_p \int_0^L dx \Psi^p \sigma_3 \partial_1 \Psi^p ,\label{H_f,diag} \\
H_{\rm f,off} 
&\!\!\!
=
&\!\!\!
 \sum_{i<j} \int_0^L dx \varphi_{ij}^\dagger 
\sigma_3 \left( \frac{1}{i} \partial_1 - g(a_i - a_j) \right) 
\varphi_{ij} ,\label{H_f,off} \\
H_{\rm b,diag} 
&\!\!\!
=
&\!\!\!
 \sum_p \int_0^L dx
\left( \frac{1}{2} \pi^p \pi^p + 
\frac{1}{2} \left(\partial_1 \phi^p \right) 
\left( \partial_1 \phi^p \right) \right),\label{H_b,diag} \\
H_{\rm b,off} &=& \sum_{i<j} \int_0^L dx \Bigl\{
\eta_{ij}^\dagger \eta_{ij} 
{}+ \left(\partial_1 \xi_{ij}^\dagger 
- i g (a_j - a_i) \xi_{ij}^\dagger \right)
\left(\partial_1 \xi_{ij}
- i g (a_i - a_j) \xi_{ij} \right) \Bigr\}.\label{H_b,off}
\end{eqnarray}


%

Let us now discuss the range 
of the variables $a^p$ \cite{Lenz}. 
Eq.(\ref{a^p def}) shows that 
the $g L a$ are angular variables
which are defined only in modulo $2 \pi$.
If the parameterization of $a$ is one-to-one 
and permutations of the eigenvalues are contained in a single domain, 
the domain is called {\it the elementary cell}. 
{}For example, in the SU(2) case, 
two eigenvalues of the matrix $a$ are 
$a_1 = a^3/2$ and $a_2 = -a^3/2$. 
Then, the elementary cell is the interval 
$
- \pi \le {gLa^3 \over 2} \le \pi,
$
with the end points identified. 
If $a^3$ is negative in the elementary cell, 
the Weyl reflection $a^3 \to -a^3$ 
maps the interval 
$-{2\pi \over gL}<a^3<0$ onto the interval $0<a^3<{2\pi \over gL}$ 
(simultaneously, 
$\varphi^{12}\leftrightarrow\varphi^{21}$). 
In the $SU(N)$ case, similarly, 
the elementary cell is divided into $N!$ domains by 
the Weyl group since the Weyl group 
of $SU(N)$ is the permutation group $P_N$. 
These $N!$ domains are called {\it fundamental domains}. 
Boundaries of the fundamental domains consist of the hypersurfaces 
where two of the eigenvalues match. 
If two of the eigenvalues have the same value, the Jacobian 
$J[a]$ vanishes. 
In the case of $SU(2)$, we take the following interval 
as the fundamental region 
\begin{eqnarray}
0 \le a^{3} \le \frac{2\pi}{gL} .
\end{eqnarray}
The Jacobian $J[a]={\rm sin}^2 \left({1 \over 2}gLa^3 \right)$ 
vanishes at $a^3=0,~\frac{2\pi}{gL}$. 
Note that the modified wavefunction $\tilde{\Phi}[a]$ vanishes
at these points.
%
%
%
\vspace{7mm}
\pagebreak[3]
\addtocounter{section}{1}
\setcounter{equation}{0}
\setcounter{subsection}{0}
\addtocounter{footnote}{0}
\begin{center}
{\large {\bf \thesection. Vacuum Structures of SUSY $SU(2)$ Yang-Mills 
Theories 
}}
\end{center}
\nopagebreak
\medskip
\nopagebreak
\hspace{3mm}

In this section, we determine the wave function of the vacuum state 
in the fundamental domain 
by using the Born-Oppenheimer approximation \cite{Lenz}. 
If $gL \ll 1$, the energy scale of the system of $a^p$ is given 
by $(gL)^2/L$, while that of 
non-zero modes of
$\Psi$ and $\phi$ is in general of order 
$1/L$.
Therefore we can integrate the non-zero modes of $\Psi$ and $\phi$
to obtain the effective potential for $a^p$. 
We will retain the zero modes of $\Psi$ and $\phi$,
since their spectrum is continuous. 
By solving the Schr\"odinger equation with respect to the 
resulting effective potential, 
we obtain the wavefunction $\tilde\Phi[a]$, which describes the 
vacuum structures of our model. 
In these procedures we must pay attention to the boundary conditions 
for $\tilde\Phi[a]$ resulting from the Jacobian 
(\ref{boundary}). 

To calculate the effective potential as a function of 
the gauge zero modes $a^p$, 
we have to find the ground state of fermion $\Psi$ and boson $\phi$ 
for a fixed value of $a^p$. 
Here, we must take care with regards to the boundary conditions 
for $\Psi(x)$ and 
$\phi(x)$. 
Since spinors and scalars are superpartners of gauge fields 
which obey the periodic boundary condition, the spinors $\Psi(x)$ and 
scalars $\phi(x)$ should be periodic in order for the boundary 
conditions to maintain supersymmetry 
\begin{eqnarray}
\Psi(x=0) = \Psi(x=L) , \qquad 
\phi(x=0) = \phi(x=L).
\end{eqnarray}
Hereafter we refer to this boundary condition as {\it the (P,P) case}.
In this section we investigate the vacuum structures
for the  
gauge group $SU(2)$. 
We will discuss other boundary conditions later. 


%
\vspace{7mm}
\pagebreak[3]
\addtocounter{section}{0}
\addtocounter{equation}{0}
\addtocounter{subsection}{1}
\addtocounter{footnote}{0}
\begin{center}
{\large {\bf \thesubsection. Born-Oppenheimer Approximation}}
\end{center}
\nopagebreak
\medskip
\nopagebreak
\hspace{3mm}

{}For $gL \ll 1$, 
the Coulomb energy (\ref{H_c}) 
and the Yukawa interaction (\ref{H_int}) can be neglected.
In this limit, the relevant parts of the Hamiltonian are, for $SU(2)$, 
\begin{eqnarray}
\tilde{H}
&\!\!\!
 =
&\!\!\!
 K_a^\prime + H_{\rm b,diag} + H_{\rm b,off}
+ H_{\rm f,diag} + H_{\rm f,off} .\label{tilde H} \\
K_a^\prime 
&\!\!\!
=
&\!\!\!
 -\frac{1}{2L}
\frac{\partial^2}{\partial a^2} + V^{[N=2]}, \\
H_{\rm b,diag} 
&\!\!\!
=
&\!\!\!
 \frac{1}{2} \int_0^L dx 
\left\{ \pi^3 \pi^3 +(\partial_1 \phi^3)(\partial_1 \phi^3)
 \right\}\label{b,diag} \\
H_{\rm b,off} 
&\!\!\!
=
&\!\!\!
 \int_0^L dx
\left\{ \eta^\dagger \eta + (\partial_1 \xi^\dagger +iga\xi^\dagger)
(\partial_1 \xi - iga\xi) \right\} ,\label{b,off} \\
H_{\rm f,diag} 
&\!\!\!
=
&\!\!\!
 \frac{1}{2i} \int_0^L dx
\Psi^3 \sigma_3 \partial_1 \Psi^3, \label{f,diag}
\qquad
H_{\rm f,off} 
=
 \int_0^L dx
\varphi^\dagger \sigma_3 \left( \frac{1}{i}
\partial_1
 -ga \right) \varphi, \label{f,off} \\
 \varphi 
&\!\!\!
\equiv
&\!\!\!
 \varphi_{12},\quad
\xi \equiv \xi_{12},\quad
\eta \equiv \eta_{12}, \quad
a \equiv a^3 =a_1 - a_2 . 
\end{eqnarray}

A remnant of large gauge transformations becomes a discrete symmetry 
$S$ \cite{Lenz} 
\begin{eqnarray}
S:
&\!\!\!
&\!\!\!
\quad a \to -a+\frac{2 \pi}{gL} , \nonumber \\
\varphi 
&\!\!\!
\to
&\!\!\!
 e^{2i \pi x/L} \varphi^{\dagger} \ ,
\quad
\xi \to e^{2i \pi x/L} \xi^\dagger \ ,
\quad
\eta \to e^{2i \pi x/L} \eta^\dagger \ ,
\nonumber \\
\Psi^3 
&\!\!\!
\to 
&\!\!\!
-\Psi^3, 
\quad
\phi^3 \to -\phi^3, 
\quad
\pi^3 \to -\pi^3.
\label{S}
\end{eqnarray}
This operator can be chosen to satisfy 
$S^2 =1$ and $[S,H]=0$. 
SYM$_2$ 
has a topologically nontrivial structure $\pi_1[SU(N)/Z_N]=Z_N$. 
The symmetry $S$ corresponds to a nontrivial element of 
this $Z_{N=2}$ group for $SU(2)$. 


In order to perform the Born-Oppenheimer approximation, we first expand 
the spinor fields $\varphi$ and $\Psi^3$, and impose a canonical 
anticommutation relation 
\begin{eqnarray}
\varphi\left(x\right)
&\!\!\!
=
&\!\!\!
\frac{1}{\sqrt{L}}\sum_{k=-\infty}^\infty{a_k \choose b_k}
e^{i 2 \pi k x /L} , 
\qquad
\left\{ a_k,a_{k^\prime}^\dagger \right\} =
\left\{ b_k,b_{k^\prime}^\dagger \right\} =
\delta_{k,k^{\prime}} ,
\nonumber \\
\Psi^3\left(x\right)
&\!\!\!
=
&\!\!\!
\frac{1}{\sqrt{L}}\sum_{k=-\infty}^\infty{c_k \choose d_k}
e^{i 2 \pi k x /L} ,
\qquad c_{-k}=c_k^\dagger ,
\quad d_{-k}=d_k^\dagger ,
\label{expand;fermion} \\
&\!\!\!
&\!\!\!
\left\{ c_k,c_{k^\prime}^\dagger \right\} =
\left\{ d_k,d_{k^\prime}^\dagger \right\} =
\delta_{k,k^{\prime}},\quad
k, k^{\prime} \ge 0 \nonumber
\end{eqnarray}
The Hamiltonian $H_{\rm f,off}$ in (\ref{f,off}) takes the form 
\begin{eqnarray}
H_{\rm f,off} = \sum_{k =-\infty}^{\infty}
\left(a_k^\dagger a_k - b_k^\dagger b_k \right)
\left(\frac{2 \pi k}{L} - ga \right) .
\end{eqnarray}
In the Born-Oppenheimer approximation,
the vacuum state for the off-diagonal part of the fermion
is obtained by filling the Dirac sea for the fermion $\varphi$. 
We assume the $a_k$ modes to be filled for $k<M$. 
The Gauss law constraint (\ref{Gauss cond.}) dictates that the 
$b_k$ modes should be filled for $k \ge M$ \cite{Lenz}.
Denoting the vacuum state for the fermion as 
$| 0_\varphi ;M \rangle$, the vacuum energy can be written as 
\begin{eqnarray}
H_{\rm f,off} | 0_\varphi ;M \rangle &=& \left[
\sum_{k=-\infty}^{M-1} \left(\frac{2 \pi k}{L} -ga \right)
- \sum_{k=M}^{\infty} \left(\frac{2 \pi k}{L} -ga \right)
\right] | 0_\varphi ;M \rangle \nonumber \\
&\equiv& V_{\rm f,off}(a;M)| 0_\varphi ;M \rangle .
\label{ev of f,off}
\end{eqnarray}
Notice that $S$ acts on the state $| 0_\varphi ;M \rangle$ 
according to 
\begin{eqnarray}
S |0_\varphi ;M \rangle
= e^{i \alpha_M} | 0_\varphi ; 2-M \rangle .
\label{actionS;pp}
\end{eqnarray}
In addition, the phase factor $e^{i \alpha_M}$ is constrained 
by $S^2=1$, or in other words, $e^{i \alpha_M} = e^{-i \alpha_{-M+2}}$. 

{}For diagonal part of the fermion, 
we obtain the Hamiltonian from (\ref{f,diag}) 
\begin{eqnarray}
H_{\rm f,diag} = \sum_{k \ge 1} \frac{2 \pi k}{L}
\left( c_k^\dagger c_k + d_k d_k^\dagger -1 \right) .
\end{eqnarray}
On the vacuum $|0_{\Psi}\rangle$ defined by 
$c_k|0_{\Psi}\rangle=d^\dagger_k|0_{\Psi}\rangle=0,~~ k\ge 1$,
we find
\begin{eqnarray}
H_{\rm f,diag}|0_{\Psi} \rangle = - \sum_{k \ge 1} \frac{2 \pi k}{L}
|0_{\Psi} \rangle 
\equiv V_{\rm f,diag} |0_{\Psi} \rangle.
\label{ev of f,diag}
\end{eqnarray}


Next we expand the scalar fields $\xi$, $\eta$, $\phi^3$, and $\pi^3$, 
and impose canonical commutation relations 
\begin{eqnarray}
\xi\left(x\right) 
&\!\!\!
=
&\!\!\!
\sum_{k=-\infty}^\infty \frac{1}{\sqrt{2L E_k}}
\left(e_k + f_k^\dagger \right)
e^{i 2 \pi k x /L} ,
\quad E_k=\left|\frac{2 \pi k}{L} -ga \right| ,\\
\eta\left(x\right) 
&\!\!\!
=
&\!\!\!
\sum_{k=-\infty}^\infty i \sqrt{\frac{E_k}{2L}}
\left(- e_k + f_k^\dagger \right)
e^{i 2 \pi k x /L} ,\\
\phi^3\left(x\right) 
&\!\!\!
=
&\!\!\!
\sum_{k=-\infty \atop k \ne 0}^\infty \frac{1}{\sqrt{2L F_k}}
\left(g_k + g_{-k}^\dagger \right)
e^{i 2 \pi k x /L} 
+ \phi_{\rm zero} ,
\quad F_k=\left|\frac{2 \pi k}{L} \right| ,\\
\pi^3\left(x\right) 
&\!\!\!
=
&\!\!\!
\sum_{k=-\infty \atop k \ne 0}^\infty i \sqrt{\frac{F_k}{2L}}
\left(- g_k + g_{-k}^\dagger \right)
e^{i 2 \pi k x /L} 
+\frac{1}{L} \pi_{\phi_{\rm zero}} .
\end{eqnarray}
\begin{eqnarray}
\left[ e_k,e_{k^\prime}^\dagger \right] =
\left[ f_k,f_{k^\prime}^\dagger \right] =
\left[ g_k,g_{k^\prime}^\dagger \right] =
\delta_{k,k^{\prime}}, \quad 
\left[ \phi_{\rm zero} , \pi_{\phi_{\rm zero}} \right] = i .
\end{eqnarray}

The Hamiltonian $H_{\rm b,off}$ in (\ref{b,off}) is given by 
\begin{eqnarray}
H_{\rm b,off}
&\!\!\!
=
&\!\!\!
 \sum_{k = -\infty}^{\infty}
E_k \left( e_k^\dagger e_k + f_k f_k^\dagger \right) \\
&\!\!\!
=
&\!\!\!
 \sum_{k = -\infty}^{\infty}
E_k \left( e_k^\dagger e_k + f_k^\dagger f_k \right) 
{}- \sum_{k=-\infty}^{N-1} \left( \frac{2 \pi k}{L} - ga \right)
+ \sum_{k=N}^{\infty} \left( \frac{2 \pi k}{L} -ga \right) ,
\end{eqnarray}
where $N$ is an integer satisfying 
\begin{equation}
\frac{2 \pi N}{L} -ga \ge 0 ,\qquad \frac{2 \pi (N-1)}{L} -ga < 0 .
\label{N}
\end{equation}
On the vacuum state $|0_{\xi}\rangle$ defined by 
$e_k |0_\xi \rangle = f_k |0_\xi \rangle = 0,~~ \mbox{for all }k$, 
we find the vacuum energy 
\begin{eqnarray}
H_{\rm b,off} | 0_\xi \rangle &=& \left[
- \sum_{k=-\infty}^{N-1} \left( \frac{2 \pi k}{L} - ga \right)
+ \sum_{k=N}^{\infty} \left( \frac{2 \pi k}{L} -ga \right)
\right] | 0_\xi \rangle \nonumber \\
&\equiv& V_{\rm b,off}(a)| 0_\xi \rangle .
\label{ev of b,off}
\end{eqnarray}

We find that the zero mode Hamiltonian $H_0$ is separated as 
\begin{eqnarray}
H_{\rm b,diag} 
&\!\!\!
= 
&\!\!\!
\sum_{k \ge 1} \frac{2 \pi k}{L}
\left( g_k^\dagger g_k + g_{-k}^\dagger g_{-k} +1 \right) +H_0 , \\
H_0 
&\!\!\!
=
&\!\!\!
 \frac{1}{2L} \pi_{\phi_{\rm zero}} \pi_{\phi_{\rm zero}} .
\label{zeromodehamiltonian}
\end{eqnarray}
On the vacuum for the nonzero modes of $\phi^3$ 
satisfying $g_k|0_{\phi} \rangle = g_{-k}|0_{\phi}\rangle =0 , k\ge 1$, 
we find the vacuum energy
\begin{eqnarray}
V_{\rm b,diag} = \sum_{k \ge 1} \frac{2 \pi k}{L}.
\end{eqnarray}

%
\vspace{7mm}
\pagebreak[3]
\addtocounter{section}{0}
\addtocounter{equation}{0}
\addtocounter{subsection}{1}
\addtocounter{footnote}{0}
\begin{center}
{\large {\bf \thesubsection. Vacuum Structure}}
\end{center}
\nopagebreak
\medskip
\nopagebreak
\hspace{3mm}

The vacuum energies obtained in the previous section 
are divergent. 
By regularizing them with
the heat kernel, 
we obtain the following finite 
effective potential as a function of $a$ 
\begin{eqnarray}
U_{M,N}(a)
&\!\!\!
=
&\!\!\!
V_{\rm f,off}(a;M) + V_{\rm b,off}(a) +
V_{\rm f,diag} + V_{\rm b,diag} 
+V^{[N=2]} 
\nonumber \\
&\!\!\!
=
&\!\!\!
\frac{2 \pi}{L} 
\left( M - \frac{gLa}{2 \pi} - \frac{1}{2} \right)^2
-\frac{2 \pi}{L} 
\left( N - \frac{gLa}{2 \pi} - \frac{1}{2} \right)^2
+V^{[N=2]} .
\label{effpotzeromode}
\end{eqnarray}
In the fundamental region $0<{gLa \over 2}< \pi$, $N=1$ 
from (\ref{N}). 
By requiring that the vacuum energy $U_{M,N}(a)$ 
be minimal,
we can fix $M$ to obtain $M=1$. 
We then find that the total vacuum energy in the 
fundamental domain is independent of $a$ 
\begin{equation}
U_{M,N}(a)=
V^{[N=2]} .
\end{equation}
Consequently we obtain the Hamiltonian which describes the vacuum 
structures for the periodic boundary condition 
\begin{equation}
\tilde{H} = K_a^\prime + H_0 
=-\frac{1}{2L}\frac{\partial}{\partial a}
\frac{\partial}{\partial a} +V^{[N=2]} +
\frac{1}{2L} \pi_{\phi_{\rm zero}} \pi_{\phi_{\rm zero}} .
\label{zeroH}
\end{equation}
We also have the zero modes of the fermion, which form a Clifford 
algebra 
\begin{eqnarray}
\Psi_{\rm zero}^3
&\!\!\!
=
&\!\!\!
 \frac{1}{\sqrt{L}}{c_0 \choose d_0},
\qquad c_0=c_0^\dagger ,
\quad d_0=d_0^\dagger ,\\
\left\{ \lambda , \lambda^\dagger \right\} 
&\!\!\!
=
&\!\!\!
 1 ,\quad
\left\{ \lambda , \lambda \right\}=
\left\{ \lambda^\dagger , \lambda^\dagger \right\}=0 ,\quad 
\lambda \equiv \frac{1}{\sqrt{2}} ( c_0 + i d_0 ) . 
\label{clifford;algebra}
\end{eqnarray}
Let us now solve the Schr\"odinger equation 
\begin{equation}
\tilde{H}\tilde{\Phi}(a)=e\tilde{\Phi}(a).
\end{equation}
Because of the boundary condition (\ref{boundary}) 
we get the wavefunction $\tilde \Phi(a)$ and the energy eigenvalue $e$ 
of the ground state as 
\begin{equation}
\tilde{\Phi}(a) = \sqrt{\frac{gL}{\pi}} 
\sin \left( \frac{gLa}{2} \right),
\qquad 
e=0.
\label{the ground state and e}
\end{equation}
It is interesting to note that the vacuum energy 
associated with the nontrivial zero mode wavefunction
(\ref{the ground state and e})
cancels precisely the contribution $V^{[N=2]}$ from the 
Jacobian in (\ref{V}).
Therefore we have shown explicitly that the SUSY is not broken 
spontaneously. 
Also note that our result is consistent with the previous 
calculation of the nonvanishing Witten index \cite{Li}. 
The calculation, however, ignores the Jacobian (\ref{J}), which is 
an important ingredient in our present attempt to define 
the gauge field zero modes properly \cite{Lenz}. 
Therefore the above explicit demonstration of the vanishing vacuum
energy using the Born-Oppenheimer approximation can be regarded as
another independent proof of 
the unbroken SUSY in SUSY 
Yang-Mills theories in $1+1$ dimensions. 


We define the vacuum 
state of the zero modes of the fermion $c_0, d_0$. Note that the zero 
modes belong to the two-dimensional representation of the Clifford 
algebra (\ref{clifford;algebra}). 
We define $| \Omega \rangle$ to be the Clifford vacuum 
annihilated by $\lambda$ and $| \tilde{\Omega} \rangle =
\lambda^\dagger | \Omega \rangle$.
Since the field $\phi^3$ can take unbounded values,
the zero mode spectrum is continuous.
This fact makes the Witten index ill-defined.
The previous attempt to compute the Witten index employed
a regularization by putting a cut-off on the $\phi_{\rm zero}$ space.
In that case, the Witten index can be defined and obtains
${\rm tr}(-1)^{F} = 1$ \cite{Li}.
In spite of this complication, we can choose the wave function
to be constant in the $\phi^3$ zero mode as the vacuum:
$H_0 | \omega \rangle =0$.

Let us now examine the transformation property under the 
discrete gauge transformation $S$.
The non-zero mode vacuum $|0_\varphi; M=1 \rangle$
turns out to be an eigenstate of $S$
\begin{equation}
S|0_\varphi; M=1 \rangle = \pm|0_\varphi; M=1 \rangle
\end{equation}
because of eq.(\ref{actionS;pp}) and $S^2=1$.
Similarly $|0_\Psi \rangle$, $|0_\xi \rangle$,
$|0_\phi \rangle$ and $| \omega \rangle$
are eigenstate of $S$ with eigenvalues $\pm 1$.
{}For the fermion zero mode, $S| \Omega \rangle = \pm |\Omega \rangle$
and  $S| \tilde{\Omega} \rangle = \mp | \tilde{\Omega} \rangle$.
Since we should construct the full vacuum state as an eigenstate with
eigenvalue
$\pm 1$ for $S$
\begin{eqnarray}
|{\bf 0}_{\Omega} \rangle 
&\!\!\! \equiv &\!\!\! 
|\tilde{\Phi}(a) \rangle
|0_\varphi ;M=1 \rangle
|0_\Psi \rangle |0_\xi \rangle |0_\phi \rangle
| \omega \rangle
| \Omega \rangle , \nonumber \\
|{\bf 0}_{\tilde{\Omega}} \rangle 
&\!\!\! \equiv &\!\!\! 
|\tilde{\Phi}(a) \rangle
|0_\varphi ;M=1 \rangle
|0_\Psi \rangle |0_\xi \rangle |0_\phi \rangle
| \omega \rangle
| \tilde{\Omega} \rangle .
\end{eqnarray} 
We find the vacuum condensate
$\left| \langle {\bf 0} |  \bar{\Psi}^a \Psi^a 
| {\bf 0}  \rangle \right|= \frac{1}{L}$ for
both $| {\bf 0} \rangle = |{\bf 0}_{\Omega} \rangle$ and
$|{\bf 0}_{\tilde{\Omega}} \rangle$.
One can see that this condensate is due to the finite spacial extent $L$.

%
%
%
\vspace{7mm}
\pagebreak[3]
\addtocounter{section}{1}
\setcounter{equation}{0}
\setcounter{subsection}{0}
\addtocounter{footnote}{0}
\begin{center}
{\large {\bf \thesection. Cases with Other Boundary Conditions}}
\end{center}
\nopagebreak
\medskip
\nopagebreak
\hspace{3mm}


In this section, we study other boundary conditions for 
the fermions $\Psi$ and the bosons $\phi$. 
There are four cases, depending on the 
choice of periodic or antiperiodic 
boundary conditions 
\begin{center}
 \begin{tabular}{lll}
                       & Fermion b. c. & Boson b. c. \\
  (P,P) case (SUSY)& periodic      & periodic     \\
  (A,P) case       & antiperiodic  & periodic     \\
  (A,A) case       & antiperiodic  & antiperiodic \\
  (P,A) case       & periodic      & antiperiodic 
 \end{tabular}
\end{center}

We shall study in turn the (A, P), (A, A), and (P, A) cases, and will 
find that the vacuum energy does not vanish. 
This indicates that SUSY is broken by boundary conditions in these three 
cases. 

%
%
\vspace{7mm}
\pagebreak[3]
\addtocounter{section}{0}
\addtocounter{equation}{0}
\addtocounter{subsection}{1}
\addtocounter{footnote}{0}
\begin{center}
{\large {\bf \thesubsection. The (A, P) Case; AntiPeriodic Fermion and 
Periodic Boson}}
\end{center}
\nopagebreak
\medskip
\nopagebreak
\hspace{3mm}

{}We first discuss the (A, P) case, where the following boundary 
conditions are imposed on the fermions $\Psi$ and bosons $\phi$ 
\begin{eqnarray}
\Psi^a(x=0)=-\Psi^a(x=L),\quad \phi^a(x=0)=\phi^a(x=L).
\end{eqnarray}
One of the motivations for considering this case is that one can naturally 
regard $L$ as the inverse temperature in the finite-temperature 
situation. 

Let us first consider the fermionic parts of the Hamiltonian 
$H_{\rm f}$. 
We expand the spinor fields $\varphi$ and $\Psi^3$ into modes, and 
obtain the Hamiltonian $H_{\rm f,off}$ in (\ref{f,off}) 
\begin{eqnarray}
\varphi\left(x\right)
&\!\!\!
=
&\!\!\!
 \frac{1}{\sqrt{L}}\sum_{k \in {\bf Z}}{a_k \choose b_k}
e^{i 2 \pi (k+1/2) x /L} ,\\
H_{\rm f,off} 
&\!\!\!
=
&\!\!\!
 \sum_{k \in {\bf Z}}
\left(a_k^\dagger a_k - b_k^\dagger b_k \right)
\left(\frac{2 \pi \left(k + \frac{1}{2} \right)}{L} - ga \right) .
\end{eqnarray}
Similarly to the (P, P) case,
the vacuum state for the off-diagonal part of the fermion
is obtained by filling the fermion negative energy states.
We assume the $a_k$ modes to be filled for $k<M$ and the $b_k$
modes for $k \ge M$.
Then the vacuum energy of $H_{\rm f,off}$ is given by 
\begin{eqnarray}
H_{\rm f,off} | 0_\varphi ;M \rangle 
&\!\!\!
=
&\!\!\!
 \Biggl[
\sum_{k=-\infty}^{M-1}
\left( \frac{2 \pi \left( k + \frac{1}{2} \right)}{L} -ga \right)
- \sum_{k=M}^{\infty}
\left(\frac{2 \pi \left(k + \frac{1}{2} \right)}{L} -ga \right)
\Biggr] | 0_\varphi ;M \rangle \nonumber \\
&\!\!\!
\equiv
&\!\!\!
 V_{\rm f,off}(a;M)| 0_\varphi ;M \rangle .
\label{AV_f,off}
\end{eqnarray}
The symmetry operator $S$ is defined as follows
\begin{eqnarray}
S |0_\varphi ;M \rangle
= e^{i \alpha_M} | 0_\varphi ; 1-M \rangle , \qquad 
e^{i \alpha_M} = e^{-i \alpha_{-M+1}} .
\label{S;aa}
\end{eqnarray}
As for $H_{\rm f,diag}$ in (\ref{f,diag}), we obtain 
\begin{eqnarray}
\Psi^3\left(x\right)
&\!\!\!
=
&\!\!\!
 \frac{1}{\sqrt{L}}\sum_{k=-\infty}^\infty{c_k \choose d_k}
e^{i 2 \pi (k+1/2) x /L} ,
\quad c_{-k-1}=c_k^\dagger ,
\quad d_{-k-1}=d_k^\dagger. \\
H_{\rm f,diag}
&\!\!\!
 =
&\!\!\!
 \sum_{k \ge 0}
\frac{2 \pi \left(k + \frac{1}{2} \right)}{L}
\left( c_k^\dagger c_k + d_k d_k^\dagger -1 \right) .
\end{eqnarray}
Therefore the vacuum state satisfying 
$c_k|0_{\Psi}\rangle =d^\dagger_k|0_{\Psi}\rangle =0,\  (k \ge 0)$ 
has energy 
\begin{eqnarray}
H_{\rm f,diag}|0_{\Psi} \rangle = - \sum_{k \ge 0}
\frac{2 \pi \left(k +\frac{1}{2} \right)}{L}
|0_{\Psi} \rangle 
\equiv V_{\rm f,diag} |0_{\Psi} \rangle .
\end{eqnarray}

As for the bosonic part of the Hamiltonian $H_{\rm b}$ in the (A, P) 
case, we may use the result of the (P, P) case because the boundary 
conditions for the scalar fields are the same. 

\begin{figure}
 \leavevmode
 \epsfysize=10cm
 \centerline{\epsfbox{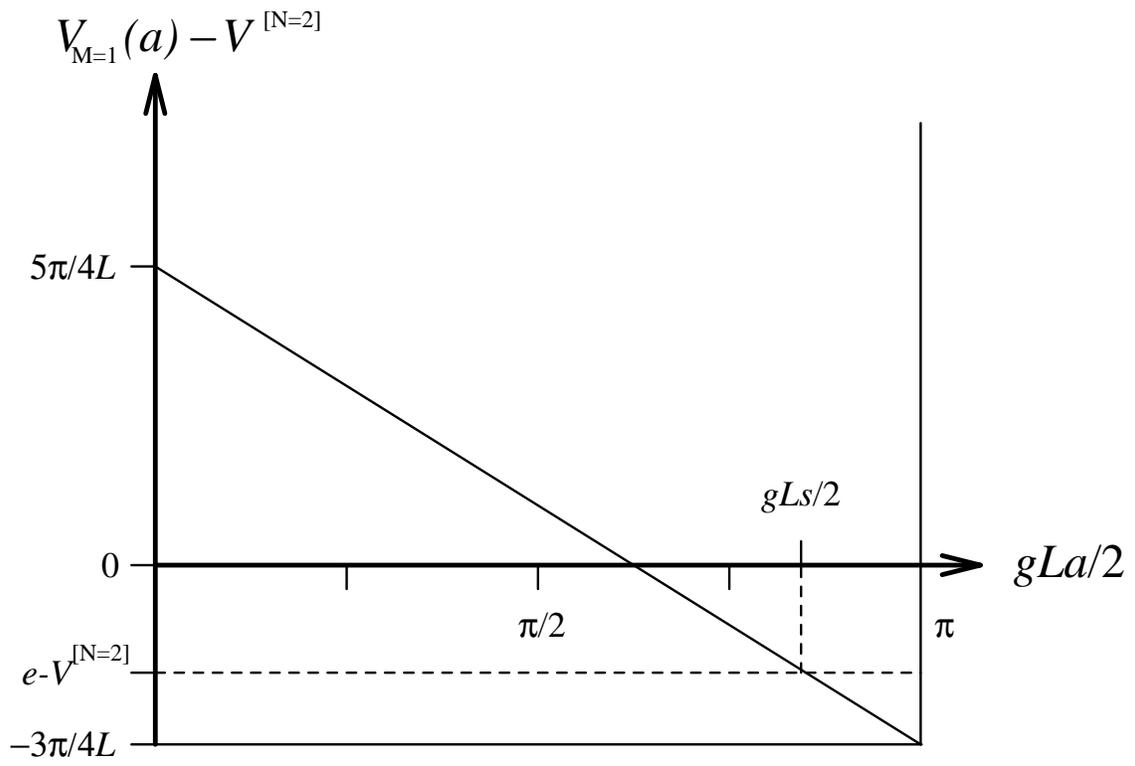}}
 \caption{
Effective potential for the (A, P) case with $M=1$ .
}
 \label{fig:APV}
\end{figure}

The heat kernel regularized potential (\ref{effpotzeromode}) 
for the gauge field zero mode $a$ becomes in this case 
\begin{equation}
U_{M, N}(a)={2\pi \over L}\left(M-{gLa \over 2\pi}\right)^2
-{2\pi \over L}\left(N-{1 \over 2}-{gLa \over 2\pi}\right)^2
-{\pi \over 4L}
+V^{[N=2]} .
\label{Umn;ap}
\end{equation}
Here, as in the (P, P) case, $N=1$. 
We can think of this as a potential in quantum mechanics for 
the zero mode $a$. 
Note that $M$ should be chosen by requiring that the ground state 
energy of the (A, P) case be minimal. 
Eq.(\ref{Umn;ap}) gives two solutions; $M=0, 1$. 
Let us first consider zero mode quantum mechanics 
in the $M=1$ case (see Fig. \ref{fig:APV}) 
\begin{eqnarray}
\tilde{H}\tilde{\Phi}_{{\rm II}}
=
e\tilde{\Phi}_{{\rm II}}, \qquad
\tilde{H} 
= 
-\frac{1}{2L} \frac{\partial^2}{\partial a^2}
+ V_{M=1}(a) +H_0.
\end{eqnarray}
\begin{eqnarray}
V_{M=1}(a)=\left\{
\begin{array}{@{\,}ll}
\infty,                & {\rm when}~~a = 0  \\
U_{1,1}(a)=-ga + \frac{5 \pi}{4L}+V^{[N=2]}, 
& {\rm when}~~0 < a < 2 \pi/gL \\
\infty,                & {\rm when}~~a = 2 \pi /gL 
\end{array}
\right.
\end{eqnarray}
The Hamiltonian for the zero mode of the scalar field 
is $H_0$ in (\ref{zeromodehamiltonian}). 


We find eigenfunctions 
with a normalization factor $
A_+$ and $A_-$ 
\begin{eqnarray}
\tilde{\Phi}_{\rm II}(a) 
&\!\!\!
=
&\!\!\!
 \left\{
\begin{array}{@{\,}ll}
\xi_2^\frac{1}{3}(a)
\left\{-
A_+ 
I_{\frac{1}{3}}\left(\xi_2(a)\right)
+ 
A_- 
I_{-\frac{1}{3}}\left(\xi_2(a)\right) \right\},
& {\rm when}~~0 \le a \le s \\
 \xi_1^\frac{1}{3}(a) 
\left\{
A_+ 
J_{\frac{1}{3}}\left(\xi_1(a)\right)
+ 
A_- 
J_{-\frac{1}{3}}\left(\xi_1(a)\right) \right\},
& {\rm when}~~s < a \le 2\pi/gL 
\end{array}
\right.
\\
\xi_1(a) 
&\!\!\!
\equiv
&\!\!\!
 \frac{2}{3} \sqrt{2gL}
\left(a - s \right)^{\frac{3}{2}} ,\quad 
\xi_2(a) \equiv \frac{2}{3} \sqrt{2gL}
\left(-a + s \right)^{\frac{3}{2}} , \\
s 
&\!\!\!
=
&\!\!\!
- \frac{1}{g} \left(e - \frac{5 \pi}{4L} -V^{[N=2]} \right),
\end{eqnarray}
where $J_\nu$ are Bessel functions and $I_\nu$ are modified Bessel 
functions.
Imposing the boundary conditions
(\ref{boundary})
on the wavefunction, 
we obtain the eigenvalues of $\tilde{H}$ as 
\begin{equation}
e_n
=
 -{3\pi \over 4L} + \frac{1}{2L}
\left( 3 b_n g L \right)^{\frac{2}{3}}
-\frac{1}{8L}(gL)^2
,\qquad
n \in {\bf N},
\end{equation}
where the $b_n$ are defined as 
\begin{equation}
I_{-{1 \over 3}}(\xi_1(0))
J_{\frac{1}{3}}(b_n) + 
I_{{1 \over 3}}(\xi_1(0))
J_{-\frac{1}{3}}(b_n)=0, \qquad b_n > 0 .
\end{equation}
Hence the ground state energy does not vanish. 
In the $M=0$ case, on the other hand, the potential is given by 
$V_{M=0}(a)=U_{\{M=0,~N=1\}}(a)=U_{\{M=1,~N=1\}}(2\pi/gL-a)$. 
It then follows that the modified wavefunction $\tilde{\Phi}_{\rm I}(a)$ 
in the $M=0$ case is 
\begin{equation}
\tilde{\Phi}_{\rm I}(a)= \tilde{\Phi}_{\rm II}(2\pi/gL-a).
\end{equation}

As discussed in \cite{Lenz}, the full vacuum state is determined by 
requiring that it should be an eigenstate of 
the symmetry $S$, which acts as 
\begin{eqnarray}
S|0_\varphi ;M \rangle = e^{i \alpha_M} |0_\varphi ;1-M \rangle ,
\qquad 
S \tilde{\Phi}_{\rm I}(a) = \tilde{\Phi}_{\rm II}(a).
\end{eqnarray}
It follows from this that the full vacuum state 
$|{\bf 0_{\pm}} \rangle $ 
is given by superposing the two states $M=0, 1$ 
\begin{eqnarray}
|{\bf 0}_{\pm} \rangle &\equiv&
\frac{1}{\sqrt{2}} \left(
|\tilde{\Phi}_{\rm I}(a) \rangle |0_\varphi ;M=0 \rangle
\pm e^{i \alpha_0}
|\tilde{\Phi}_{\rm II}(a) \rangle |0_\varphi ;M=1 \rangle
\right) \nonumber \\
& &{} \otimes |0_\Psi \rangle |0_\xi \rangle |0_\phi \rangle
|\omega \rangle .
\end{eqnarray}
$|{\bf 0}_{\pm}\rangle$ have eigenvalues $\pm 1$ of $S$. 
Here we assume that 
$|0_\Psi \rangle |0_\xi \rangle |0_\phi \rangle |\omega \rangle$
is invariant under $S$.

Let us consider the gaugino bilinear condensate in this vacuum 
$|{\bf 0_{\pm}} \rangle $. 
\begin{eqnarray}
\langle {\bf 0}_\pm | \bar{\Psi}^a \Psi^a| {\bf 0}_\pm \rangle 
= \pm \frac{2}{L} \sin \alpha_0 \langle \tilde{\Phi}_{\rm I}
| \tilde{\Phi}_{\rm II} \rangle .
\end{eqnarray}
Taking the massless limit from an infinitesimally massive
$\Psi^a$, one obtains $\alpha_0 = \frac{\pi}{2}$ \cite{LeNaTh}.
The overlap of the two wavefunctions is found as 
\begin{eqnarray}
\langle \tilde{\Phi}_{\rm I} | \tilde{\Phi}_{\rm II} \rangle 
&\!\!\!
=
&\!\!\!
\int_0^{2 \pi/gL} da
\tilde{\Phi}_{\rm I}(a) \tilde{\Phi}_{\rm II}(a) \nonumber \\
&\!\!\!
=
&\!\!\!
2^{1 \over 12}3^{2 \over 3}\pi^{1 \over 4}
{ 1 \over 
\Gamma({1 \over 3})+{2^{4/3}\pi \over 3}I}
(gL)^{1 \over 6} e^{- \frac{2(2 \pi)^{3/2}}{3gL}}
\\
I
&\!\!\!
=
&\!\!\!
\int_0^{b_1}d \xi_1~\xi_1^{{1 \over 3}}\left[J_{1 \over 3}(\xi_1)+
J_{-{1 \over 3}}(\xi)\right]^2.
\end{eqnarray}
The dependence of the vacuum condensate on 
the gauge coupling constant recalls, interestingly, the case of 
instanton contributions. 
This suggests that our result may also be derived by means of 
instanton calculus. 
The instanton-like result is a characteristic feature of this 
boundary condition. 
An interesting feature of our SUSY model compared with the non-SUSY 
models of \cite{Smilga} and \cite{Lenz} is the prefactor 
$(gL)^{1/6}$, which has a positive fractional exponent. 

%
\vspace{7mm}
\pagebreak[3]
\addtocounter{section}{0}
\addtocounter{equation}{0}
\addtocounter{subsection}{1}
\addtocounter{footnote}{0}
\begin{center}
{\large {\bf \thesubsection. The (A, A) Case; AntiPeriodic Fermion and 
Boson}}
\end{center}
\nopagebreak
\medskip
\nopagebreak
\hspace{3mm}

We impose the following boundary conditions on the fermions $\Psi$ and 
the bosons $\phi$
\begin{eqnarray}
\Psi^a(x=0) = - \Psi^a(x=L), \quad
\phi^a(x=0) = - \phi^a(x=L).
\end{eqnarray}

Obviously, the fermionic part of the Hamiltonian $H_{\rm f}$ is the same as 
that of the (A, P) case. As for $H_{\rm b}$, the derivation is 
summarized in Appendix. 
%
%
Similarly to (\ref{Umn;ap}), 
the heat kernel regularization gives the effective potential 
as a function of the zero mode $a$ of the gauge fields 
\begin{equation}
U_{M,N}(a)={2\pi \over L}\left( M-{gLa \over 2\pi} \right)^2-
{2\pi \over L}\left( N-{gLa \over 2\pi} \right)^2
+V^{[N=2]}.
\label{u_mn;ppmodel}
\end{equation}
In the fundamental domain, $N$ is given by 
\begin{eqnarray}
N
=\left\{
\begin{array}{@{\,}ll}
0, &  {\rm when}~~~0<{gLa \over 2}< {\pi \over 2}, \\
1, &  {\rm when}~~~{\pi \over 2} < {gLa \over 2}< \pi.
\end{array}
\right.
\label{N;aa}
\end{eqnarray}

\begin{figure}
 \leavevmode
 \epsfysize=10cm
 \centerline{\epsfbox{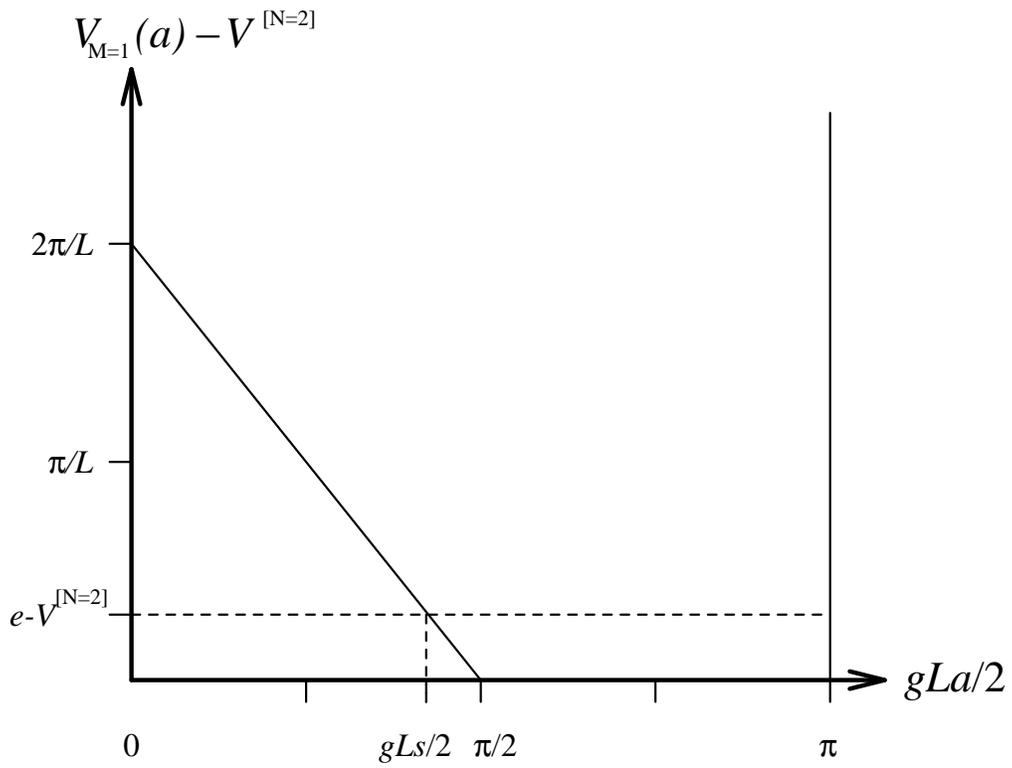}}
 \caption{
Effective potential for the (A, A) case with $M=1$ .
}
 \label{fig:AAV}
\end{figure}

Now let us discuss quantum mechanics for the zero mode $a$. 
The vacuum state of the (A, A) case is given 
by superposing the two possible states $M=0, 1$.  
For $M=1$,  we obtain (see Fig. \ref{fig:AAV}) 
\begin{equation}
\tilde{H}\tilde{\Phi}_{{\rm II}}(a)=e\tilde{\Phi}_{{\rm II}}(a), 
\qquad 
\tilde{H} 
= -\frac{1}{2L} \frac{\partial^2}{\partial a^2}
+ V_{M=1}(a) ,
\label{Schrodinger;aa}
\end{equation}
\begin{eqnarray}
V_{M=1}(a)=\left\{
\begin{array}{@{\,}ll}
\infty,                & {\rm when}~~a = 0,  \\
U_{1,0}(a)=\frac{2}{L}(\pi -gLa)+V^{[N=2]}, 
& {\rm when}~~0 < a < \pi/gL, \\
U_{1,1}(a)=V^{[N=2]},           &{\rm when}~~\pi/gL \le a <2 \pi/gL, \\
\infty,                & {\rm when}~~a = 2 \pi /gL. 
\end{array}
\right.
\end{eqnarray}
Imposing the boundary conditions (\ref{boundary})
on the wavefunction leads to the 
discrete energy spectrum 
\begin{eqnarray}
e_n= \frac{(gL)^2}{2L} n^2 
\left[ 1 - \frac{\Gamma({1 \over 3})}{\pi~\Gamma({2 \over 3})}
\left(\frac{2 (gL)^2}{3}\right)^{{1 \over 3}} \right]
-\frac{1}{8L}(gL)^2
,\qquad
n \in {\bf N}.
\end{eqnarray}
Thus, the ground state ($n=1$) has positive energy
\begin{eqnarray}
e_{\rm vac} = 
\frac{3}{8L}(gL)^2 \left[ 1 - \left( {2 \over 3}\right)^{{4 \over 3}}
\frac{2\Gamma({1 \over 3})}{\pi~\Gamma({2 \over 3})}
(gL)^{{2 \over 3}}\right]. 
\end{eqnarray}
In the $M=0$ case, the potential $V_{M=0}(a)$  has a symmetry 
$V_{M=0}(a) = V_{M=1}(2 \pi /gL -a)$. Hence the modified wavefunction in the 
$M=0$ case $\tilde{\Phi}_{\rm I}(a)$ is given by 
\begin{equation}
\tilde{\Phi}_{\rm I}(a) = \tilde{\Phi}_{\rm II}(2\pi/gL -a).
\end{equation}
Note that $S$ defined in (\ref{S;aa}) also provides a transformation between 
$\tilde{\Phi}_{\rm I}(a)$ and $\tilde{\Phi}_{\rm II}(a)$. 

{}From these results we are able to write down the full vacuum state 
vector,
which we take to be an eigenstate of the symmetry operator $S$
with eigenvalue $\pm 1$:
\begin{eqnarray}
|{\bf 0}_\pm \rangle \equiv \frac{1}{\sqrt{2}} \left( 
| \tilde{\Phi}_{\rm I}(a) \rangle
|0_\varphi ;M=0 \rangle
\pm e^{i \alpha_0}
| \tilde{\Phi}_{\rm II}(a) \rangle
|0_\varphi ;M=1 \rangle
\right)
|0_\Psi \rangle |0_\xi \rangle |0_\phi \rangle,
\end{eqnarray}
where we assume that 
$|0_\Psi \rangle |0_\xi \rangle |0_\phi \rangle$
is invariant under $S$.

We find the condensate 
on this vacuum state 
$|{\bf 0}_\pm \rangle $ as 
\begin{eqnarray}
\langle {\bf 0}_\pm | \bar{\Psi}^a \Psi^a | {\bf 0}_\pm \rangle 
= 
\pm \frac{2}{L} \sin \alpha_0
\langle \tilde{\Phi}_{\rm I} | \tilde{\Phi}_{\rm II} \rangle 
= \pm {3\sqrt{2} \over 4\pi^2L}\Big(\Gamma(1/3)\Big)^3 
\sin \alpha_0~ (gL)^{2}.
\end{eqnarray}
As in the (A, P) case, we take $\alpha_0 = \frac{\pi}{2}$.
%
%
\vspace{7mm}
\pagebreak[3]
\addtocounter{section}{0}
\addtocounter{equation}{0}
\addtocounter{subsection}{1}
\addtocounter{footnote}{0}
\begin{center}
{\large {\bf \thesubsection. The (P, A) Case; Periodic Fermion and 
AntiPeriodic Boson}}
\end{center}
\nopagebreak
\medskip
\nopagebreak
\hspace{3mm}

\begin{figure}
 \leavevmode
 \epsfysize=10cm
 \centerline{\epsfbox{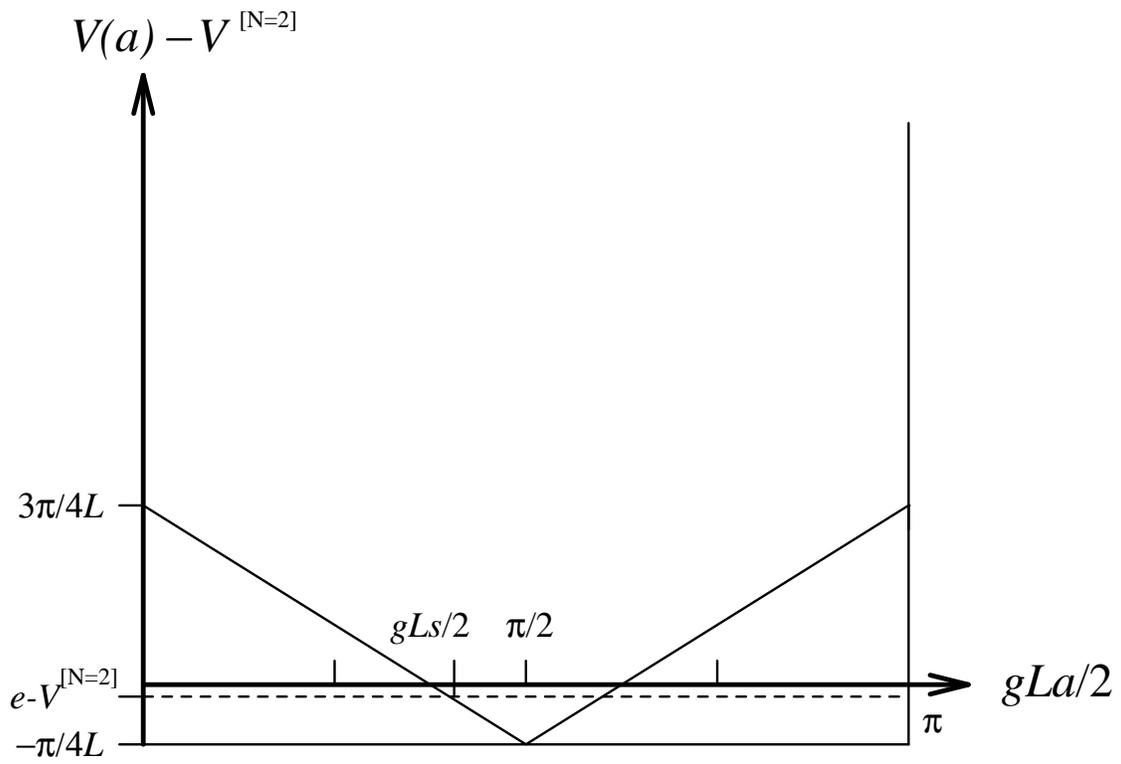}}
 \caption{
Effective potential for the (P, A) case.
}
 \label{fig:PAV}
\end{figure}

We impose 
the following boundary condition on the fermions $\Psi$ 
and bosons $\phi$ 
\begin{eqnarray}
\Psi(x=0)=\Psi(x=L),\quad \phi(x=0)=-\phi(x=L).
\end{eqnarray}
Similarly to other cases, the 
effective potential for the zero modes $a$ is found to be 
\begin{equation}
U_{M, N}(a)={2\pi \over L}\left(M-{1 \over 2}-{gLa \over 2\pi}\right)^2
-{2\pi \over L}\left(N-{gLa \over 2\pi}\right)^2+{\pi \over 4L}
+V^{[N=2]}.
\end{equation}
The integer $N$ is given by (\ref{N;aa}). 
To minimize the vacuum energy for fixed $a$, we have $M=1$. 
The effective 
potential (Fig. \ref{fig:PAV}) of the (P, A) case is given by 
\begin{eqnarray}
V(a)=\left\{
\begin{array}{@{\,}ll}
\infty,             & {\rm when}~~a = 0,  \\
U_{1,0}(a)=-ga+{3\pi \over 4L} + V^{[N=2]}, 
& {\rm when}~~0 < a < \pi/gL, \\
U_{1,1}(a)=ga-\frac{5\pi}{4L} + V^{[N=2]}, 
& {\rm when}~~\pi/gL \le a <2 \pi/gL, \\
\infty,             & {\rm when}~~a = 2 \pi /gL.
\end{array}
\right.
\end{eqnarray}

The Schr\"odinger equation takes the form 
\begin{equation}
\tilde{H}\tilde{\Phi}(a)=e\tilde{\Phi}(a), \qquad 
\tilde{H} = -\frac{1}{2L} \frac{\partial^2}{\partial a^2}
+ V(a) .
\label{Schrodinger;pa}
\end{equation}
Using $gL \ll 1$, we obtain the eigenvalues of the Hamiltonian 
\begin{eqnarray}
e_n= - \frac{\pi}{4L} + \frac{1}{2L}
\left( 3 a_n g L \right)^{\frac{2}{3}}
- \frac{1}{8L}(gL)^2
,\qquad
n \in {\bf N},
\end{eqnarray}
where the $a_n$ are defined using $\xi_2(a)$ in eq.(\ref{paxi}) of appendix  
\begin{eqnarray}
I_{-{1 \over 3}}\left(\xi_2(a=0)\right) 
J_{-\frac{2}{3}}(a_n) - 
I_{{1 \over 3}}\left(\xi_2(a=0)\right) 
J_{\frac{2}{3}}(a_n)=0, \qquad a_n > 0. 
\end{eqnarray}
Thus the ground state ($n=1$) has nonvanishing energy. 

We find that the vacuum 
state of the (P, A) case can be written as 
\begin{eqnarray}
|{\bf 0}_{\Omega} \rangle 
&\!\!\! \equiv &\!\!\! 
|\tilde{\Phi}(a) \rangle
|0_\varphi ;M=1 \rangle
|0_\Psi \rangle |0_\xi \rangle |0_\phi \rangle
| \Omega \rangle , \nonumber \\
|{\bf 0}_{\tilde{\Omega}} \rangle 
&\!\!\! \equiv &\!\!\! 
|\tilde{\Phi}(a) \rangle
|0_\varphi ;M=1 \rangle
|0_\Psi \rangle |0_\xi \rangle |0_\phi \rangle
| \tilde{\Omega} \rangle .
\end{eqnarray} 
We find the vacuum condensate
$\left| \langle {\bf 0} |  \bar{\Psi}^a \Psi^a 
| {\bf 0}  \rangle \right|= \frac{1}{L}$ for
both $| {\bf 0} \rangle = |\tilde{\Phi}(a) \rangle$ and
$|{\bf 0}_{\tilde{\Omega}} \rangle$.

%
%
%
\vspace{7mm}
\pagebreak[3]
\addtocounter{section}{1}
\setcounter{equation}{0}
\setcounter{subsection}{0}
\addtocounter{footnote}{0}
\begin{center}
{\large {\bf \thesection. Summary}}
\end{center}
\nopagebreak
\medskip
\nopagebreak
\hspace{3mm}

This paper discusses two-dimensional supersymmetric
Yang-Mills theories (SYM$_2$), which are defined on the compactified 
spatial region with interval $L$. It is possible to 
gauge away all gauge fields 
except for the zero modes. Under the condition $gL \ll 1$, 
the vacuum structures of SYM$_2$ are discussed by solving the 
quantum mechanics of the zero modes. 
The Jacobian associated with the change of variable to zero modes 
gives rise to 
nontrivial results. The vacuum states are described in terms of the 
wavefunctions depending on the zero modes. 

Depending on the choice of boundary conditions, there are four different 
cases. The first is the (P, P) case. 
The ground states of this case turn out to have vanishing energy; 
however 
we cannot count the zero energy states because of the zero modes of the 
scalar field. Such difficulties also appear in \cite{Li}. 
The gaugino bilinear condensate is calculated in the (P, P) case.
It is found that the gaugino condensate 
is independent of the gauge coupling constant $g$.

The paper also discusses three other cases 
having different boundary conditions for the spinor fields  $\Psi(x)$
and/or the scalar fields $\phi(x)$.
The ground states of these cases commonly possess nonvanishing energy. 
This suggests that these three boundary conditions 
do not preserve supersymmetry. 
Their vacuum structures are quite different from each other. 
For example, the gaugino condensates depend on the coupling
constant $g$ if and only if the boundary conditions of 
$\Psi(x)$ are antiperiodic.

Among the four cases, the one of great interest is the (A, P) case. 
The vacuum condensate 
includes nontrivial structures, which resemble those related to the 
contribution of the instantons. This similarity indicates that our results may 
also be 
obtained by using Smilga's approach \cite{Smilga}. In fact, the boundary 
conditions of the (A, P) case 
are very similar to
those of Euclidean 
field theories with finite imaginary time. Moreover, the potential 
energy of the (A, P) case exhibits a double-well structure, 
which has already been
 extensively studied from the viewpoint of instantons \cite{Pol}.

It is worth comparing the discussion of the Witten index in
2-dimentions \cite{Li}
with that in 4-dimentions \cite{Wit}. 
We notice two significant differences: 
First, 
the spectrum in the 2-dimentional case 
is continuous owing to the zero modes of the scalar field. 
Therefore the Witten index is ill-defined.
It is necessary to put a cut-off for the space of the zero modes 
in order to make the Witten index well-defined.
On the other hand,
there are no zero modes of the scalar field
in the 4-dimentional case. 
Therefore the spectrum is discrete, and the Witten index is 
well-defined. 
Second, the 4-dimentional theory has a complex Weyl spinor while 
the 2-dimensional case contains a Majorana spinor. 
In the 4-dimensional case, the complex Weyl spinor
can be written in the form 
of the creation and annihilation
operators $a_{\alpha}^{\dagger \sigma}$ and 
$a_{\alpha}^{\sigma}$ 
$(\alpha =1,2 ; \sigma = 1 , \ldots,r)$ satisfying 
$\{ a_{\alpha}^{\sigma}, a_{\beta}^{\dagger \tau} \} =
\delta_{\alpha \beta} \delta^{\sigma \tau}$,
where $r = N-1$. 
Let $| \Omega \rangle $ be the Clifford vacuum which is annihilated by 
$a_{\alpha}^{\sigma}$. Depending on whether $| \Omega \rangle $ is invariant 
or pseudo-invariant state under the Weyl group, there are two possible cases 
for the Witten index. When $| \Omega \rangle $ is invariant, the Weyl 
invariant zero energy states can be given in the form of 
$| \Omega \rangle , U|\Omega \rangle , \ldots , U^r | \Omega \rangle$, 
where $U$ is the Weyl invariant operator given by 
$U = a_{\alpha}^{\dagger \sigma} a_{\beta}^{\dagger \sigma} 
\epsilon^{\alpha \beta}$. It then follows ${\rm tr}(-1)^{F}= r+1$.
In the case where $| \Omega \rangle $ is pseudo-invariant, 
the Weyl invariant states can be given by acting on $|\Omega \rangle$
the pseudo-invariant 
operators 
$V_{\alpha_1 \cdots \alpha_r} =
a_{\alpha_1}^{\dagger \sigma_1} \cdots
a_{\alpha_r}^{\dagger \sigma_r}
\epsilon_{\sigma_1 \cdots \sigma_r}$,
which have spin $r/2$.
Thus, one obtains ${\rm tr}(-1)^{F}=(-1)^{r+1} (r+1)$. These two 
results imply that there is an ambiguity of the sign of the Witten index.
In four dimensions, this ambiguity cannot be removed. 
On the other hand, in the 2-dimensional case, there is only one ground state 
and moreover such an ambiguity of the sign does not 
appear. In the present case, the zero modes of the Majorana spinor satisfy 
$\{ \lambda^\sigma , \lambda^{\dagger \tau} \} = \delta^{\sigma \tau}$
(see (\ref{clifford;algebra})).
If the Clifford vacuum is Weyl-invariant, this is the unique zero energy 
state because there is no Weyl-invariant operator. If the Clifford vacuum 
is pseudo-invariant, the allowed zero energy state is given by acting the 
pseudo-invariant operator 
$V= \epsilon_{\sigma_1 \cdots \sigma_r} 
\lambda^{\dagger \sigma_1} \cdots \lambda^{\dagger \sigma_r}$.
Thus, in both cases, there is only one ground state as long as the 
subtleties associated with the zero modes of the scalar field are overcome. 
One can always redefine the fermionic number to change fermions
into bosons and {\it vice versa},
since the fermionic number does not have an intrinsic meaning
in two dimensions.
We have explicitly demonstrated the above mechanism
in the case of $SU(2)$.

\vspace{10mm}

We would like to thank Dr. S. Kojima for participating in the early stage of 
this work and for useful comments. We also acknowledge Christian Baraldo 
for reading of the manuscript and useful suggestions.

\def\numberbysectiona{\@addtoreset{equation}{section}
\def\theequation{A.\arabic{equation}}}
\numberbysectiona
\vspace{7mm}
\pagebreak[3]
\setcounter{section}{1}
\setcounter{equation}{0}
\setcounter{subsection}{0}
\setcounter{footnote}{0}
\begin{center}
{\large{\bf Appendix }}
\end{center}
\nopagebreak
\medskip
\nopagebreak
\hspace{3mm}

Here we discuss the vacuum states of $H_{\rm b}$ in the (A, A), and 
the (P, A) 
cases, and derive the vacuum energy of these cases. 
First we consider the (A, A) case. 
The mode expansions of the scalar fields $\xi$, $\eta$, $\phi^3$, 
and $\pi^3$ take the form 
\begin{eqnarray}
\xi\left(x\right) 
&\!\!\!
=
&\!\!\!
\sum_{k=-\infty}^\infty \frac{1}{\sqrt{2L E_k}}
\left(e_k + f_k^\dagger \right)
e^{i 2 \pi (k +1/2)x /L} , 
\quad  E_k=\left|\frac{2 \pi
\left(k + \frac{1}{2} \right)}{L} -ga \right| , \\
\eta\left(x\right) 
&\!\!\!
=
&\!\!\!
\sum_{k=-\infty}^\infty i \sqrt{\frac{E_k}{2L}}
\left(- e_k + f_k^\dagger \right)
e^{i 2 \pi (k+1/2) x /L} ,\\
\phi^3\left(x\right) 
&\!\!\!
=
&\!\!\!
\sum_{k=-\infty}^\infty \frac{1}{\sqrt{2L F_k}}
\left(g_k + g_{-k-1}^\dagger \right)
e^{i 2 \pi (k+1/2) x /L} ,
\quad F_k=\left|\frac{2 \pi
\left( k + \frac{1}{2} \right)}{L} \right| ,\\
\pi^3\left(x\right) 
&\!\!\!
=
&\!\!\!
\sum_{k=-\infty}^\infty i \sqrt{\frac{F_k}{2L}}
\left(- g_k + g_{-k-1}^\dagger \right)
e^{i 2 \pi (k+1/2) x /L} ,
\end{eqnarray}
where $e_k$, $f_k$ and $g_k$ satisfy the commutation relations, 
\begin{eqnarray}
\left[ e_k,e_{k^\prime}^\dagger \right] =
\left[ f_k,f_{k^\prime}^\dagger \right] =
\left[ g_k,g_{k^\prime}^\dagger \right] =
\delta_{k,k^{\prime}}.
\end{eqnarray}
We evaluate the Hamiltonian $H_{\rm b,off}$ in (\ref{b,off}) 
by defining an integer $N=\left[{gaL \over 2\pi}+{1 \over 2}\right]$ 
\begin{eqnarray}
H_{\rm b,off}
&\!\!\!
=
&\!\!\!
 \sum_{k = -\infty}^{\infty}
E_k \left( e_k^\dagger e_k + f_k^\dagger f_k \right) \nonumber \\
&\!\!\!
&\!\!\!
 {}- \sum_{k=-\infty}^{N-1}
\left( \frac{2 \pi \left( k + \frac{1}{2} \right)}{L} - ga \right)
+ \sum_{k=N}^{\infty}
\left( \frac{2 \pi \left( k + \frac{1}{2} \right)}{L} -ga \right) .
\end{eqnarray}
Then the vacuum state satisfying 
$e_k |0_\xi \rangle = f_k |0_\xi \rangle = 0$ 
has energy given by 
\begin{eqnarray}
H_{\rm b,off} | 0_\xi \rangle 
&\!\!\!
=
&\!\!\!
 \left[
- \sum_{k=-\infty}^{N-1}
\left( \frac{2 \pi \left(k + \frac{1}{2} \right)}{L} - ga \right)
+ \sum_{k=N}^{\infty}
\left( \frac{2 \pi \left( k + \frac{1}{2} \right)}{L} -ga \right)
\right] | 0_\xi \rangle \nonumber \\
&\!\!\!
\equiv
&\!\!\!
 V_{\rm b,off}(a)| 0_\xi \rangle .
\end{eqnarray}
We also obtain the Hamiltonian $H_{\rm b,diag}$ in (\ref{b,diag}) as 
\begin{eqnarray}
H_{\rm b,diag} = \sum_{k \ge 0}
\frac{2 \pi \left(k + \frac{1}{2} \right)}{L}
\left( g_k^\dagger g_k + g_{-k-1}^\dagger g_{-k-1} +1 \right).
\end{eqnarray}
Defining the vacuum state as 
$g_k|0_{\phi} \rangle =0$ for all $k$,
the vacuum energy is given by 
\begin{eqnarray}
V_{\rm b,diag} = \sum_{k \ge 0}
\frac{2 \pi \left( k + \frac{1}{2} \right)}{L}.
\end{eqnarray}
We find for the vacuum energy of the 
antiperiodic boson 
\begin{equation}
H_{\rm b}|0_{\xi}\rangle |0_{\phi}\rangle = 
\Big( V_{\rm b,off}(a) + V_{\rm b,diag}\Big)
|0_{\xi}\rangle |0_{\phi}\rangle.
\end{equation}

Solving the Schr\"odinger equation (\ref{Schrodinger;aa}), 
we obtain the eigenfunctions 
\begin{eqnarray}
\tilde{\Phi}_{\rm II}(a) 
&\!\!\!
=
&\!\!\!
 \left\{
\begin{array}{@{\,}ll}
 \xi_2^\frac{1}{3}(a)
\left\{-
A_+ 
I_{\frac{1}{3}}\left(\xi_2(a)\right)
+ 
A_- 
I_{-\frac{1}{3}}\left(\xi_2(a)\right) \right\},
& {\rm when}~~0 \le a \le s, \\
 \xi_1^\frac{1}{3}(a) 
\left\{
A_+ 
J_{\frac{1}{3}}\left(\xi_1(a)\right)
+ 
A_- 
J_{-\frac{1}{3}}\left(\xi_1(a)\right) \right\},
&{\rm when}~~s < a \le \pi/gL, \\
B \sin \left( \sqrt{2 L(e-V^{[N=2]})} \left(\frac{2\pi}{gL}-a\right)\right),
&{\rm when}~~\pi/gL < a \le 2\pi/gL, 
\end{array}
\right.
\\
\xi_1(a) 
&\!\!\!
\equiv
&\!\!\!
 \frac{4}{3} \sqrt{gL}
\left(a - s \right)^{\frac{3}{2}} , \quad 
\xi_2(a) \equiv \frac{4}{3} \sqrt{gL}
\left(-a + s \right)^{\frac{3}{2}} , 
\quad 
s 
=
 \frac{\pi}{gL} - \frac{1}{2g}\left(e-V^{[N=2]}\right) .
\end{eqnarray}
Energy eigenvalue condition is given by 
\begin{eqnarray}
&\!\!\!
&\!\!\!
-\cot \left({\pi \over gL}\sqrt{2L(e-V^{[N=2]})}\right) 
\\ 
&\!\!\!
=
&\!\!\!
{
I_{-\frac{1}{3}}\left(\xi_2(a=0)\right)
J_{-\frac{2}{3}}\left(\xi_1(a={\pi \over gL})\right)
-
I_{\frac{1}{3}}\left(\xi_2(a=0)\right)
J_{\frac{2}{3}}\left(\xi_1(a={\pi \over gL})\right)
 \over 
I_{-\frac{1}{3}}\left(\xi_2(a=0)\right)
J_{\frac{1}{3}}\left(\xi_1(a={\pi \over gL})\right)
+
I_{\frac{1}{3}}\left(\xi_2(a=0)\right)
J_{-\frac{1}{3}}\left(\xi_1(a={\pi \over gL})\right)
}
\nonumber 
\end{eqnarray}

For the (P, A) case, the solutions to the Schr\"odinger equation 
(\ref{Schrodinger;pa}) are given as 
\begin{eqnarray}
\tilde{\Phi}(a) = \left\{
\begin{array}{@{\,}ll}
 \xi_2^{{1 \over 3}}(a)
\left\{-
A_+ 
I_{\frac{1}{3}}\left(\xi_2(a)\right)
+ 
A_- 
I_{-\frac{1}{3}}\left(\xi_2(a)\right) \right\},
& {\rm when}~~0 \le a \le s, \\
 \xi_1^{{1 \over 3}}(a) 
\left\{
A_+ 
J_{\frac{1}{3}}\left(\xi_1(a)\right)
+ 
A_- 
J_{-\frac{1}{3}}\left(\xi_1(a)\right) \right\},
& {\rm when}~~s < a \le \pi/gL,\\
 \xi_1^{{1 \over 3}}(a^{\prime}) 
\left\{
A_+ 
J_{\frac{1}{3}}\left(\xi_1(a^{\prime})\right)
+ 
A_- 
J_{-\frac{1}{3}}\left(\xi_1(a^{\prime})\right) \right\},
&{\rm when}~~\pi/gL < a \le 2\pi/gL -s,\\
 \xi_2^{{1 \over 3}}(a^{\prime})
\left\{-
A_+ 
I_{\frac{1}{3}}\left(\xi_2(a^{\prime})\right)
+ 
A_- 
I_{-\frac{1}{3}}\left(\xi_2(a^{\prime})\right) \right\},
&{\rm when}~~2\pi/gL -s < a \le 2\pi/gL,
\end{array}
\right.
\end{eqnarray}
\begin{eqnarray}
\xi_1(a) = \frac{2}{3} \sqrt{2gL}
\left(a - s \right)^{{3 \over 2}} , 
\quad 
\xi_2(a) = \frac{2}{3} \sqrt{2gL}
\left(-a + s \right)^{{3 \over 2}} , 
\label{paxi}
\end{eqnarray}
\begin{equation}
a^{\prime}={2\pi \over gL}-a, \qquad 
s=-\frac{1}{g}\left( e - \frac{3\pi}{4L} - V^{[N=2]} \right).
\end{equation}


%
%
\section*{Figure captions}
\begin{itemize}
\item[Fig.\ 1]
Effective potential for the (A, P) case with $M=1$ .

\item[Fig.\ 2]
Effective potential for the (A, A) case with $M=1$ .

\item[Fig.\ 3]
Effective potential for the (P, A) case.

\end{itemize}

%
%
%
%
\newpage


\end{document}